\documentclass[pra,aps,twocolumn,showpacs,epsfig]{revtex4}
\usepackage{color}
\usepackage{graphicx,amssymb,epsfig,epsf,amsmath}

\begin{document}

\title{Behavior of heat capacity of an attractive Bose-Einstein Condensate 
approaching collapse} 

\author{Sanchari Goswami\footnote{e-mail: sg.phys.caluniv@gmail.com}, Tapan Kumar Das\footnote{e-mail: tkdphy@caluniv.ac.in} and Anindya Biswas\footnote{e-mail: abc.anindya@gmail.com}}

\affiliation{Department of Physics, University of Calcutta, 
92 A.P.C. Road, Kolkata 700009, India}

\begin{abstract} 
We report calculation of heat capacity of an attractive Bose-Einstein 
condensate, with the number $N$ of bosons increasing and eventually 
approaching the critical 
number $N_{cr}$ for collapse, using the correlated potential harmonics 
(CPH) method. Boson pairs interact via the realistic van der Waals 
potential. It is found that the transition temperature $T_c$ increases 
initially slowly, then rapidly as $N$ becomes closer to $N_{cr}$. 
The peak value of 
heat capacity for a fixed $N$ increases slowly with $N$, for $N$ far away from 
$N_{cr}$. But after reaching a maximum, it starts decreasing when 
$N$ approaches $N_{cr}$. The effective potential 
calculated by CPH method provides an insight into this strange behavior. 
\vskip 5pt \noindent 
\end{abstract} 
\maketitle
\section{Introduction}
Bose-Einstein condensation (BEC) is the transition process, in which a 
macroscopic 
fraction of bosons goes into the lowest energy state, as the temperature 
is lowered below a certain critical
temperature $T_c$~\cite{Huang}. It was predicted by Einstein in 1925, based on
Bose's explanation of black body radiation. A great deal of activity, both theoretical 
and experimental, has been seen in this field, since the experimental realization of 
BEC in 1995. Although a number of static, 
dynamic and thermodynamic properties have 
been studied~\cite{Dalfovo,Pethick}, not much attention has been paid to the heat capacity of 
attractive condensates. The main motivation of this 
work is to fill this gap. \\

In laboratory experiments, the condensate is trapped by a confining potential, 
usually a harmonic oscillator potential. 
An attractive condensate ({\it e.g.} $^7$Li condensate) 
has a negative value of the $s$-wave scattering length $a_s$ and collapses, 
when the number of particles $N$ in the condensate exceeds a critical 
number $N_{cr}$. On the other hand a repulsive condensate ({\it e.g.} $^{87}$Rb 
condensate) corresponds to $a_s > 0$ and is stable for any $N$, 
since repulsively interacting  bosons are contained in the 
externally applied trap. 
The situation is quite different for an attractive condensate: attractive 
bosons tend to come to the center of the trap, which is balanced only by the 
kinetic pressure, resulting in a metastable condensate. The total attraction 
increases as the number of pairs $N(N-1)/2$, while the kinetic pressure increases as $N$. 
Thus for $N$ larger than a critical value $N_{cr}$, the net attraction dominates 
and a collapse occurs. \\

In this communication, we report the calculation of heat capacity of an 
attractive condensate containing a fixed number of $^7$Li atoms, interacting 
via the realistic van der Waals potential, appropriate for the experimental 
scattering length. 
The features are markedly different from those of repulsive condensates, only 
which have so far been investigated. In a repulsive condensate, the heat 
capacity $C_N(T)$ 
for a fixed number $N$ of bosons in the trap, as also the critical 
temperature $T_c$, smoothly 
approach a constant value as $N \rightarrow \infty$~\cite{Napolitano,Biswas1}. As a function 
of $T$, the heat capacity for a given $N$ increases to a maximum $(C_N)_{max}$, 
then falls rapidly to a saturation value $3Nk_B$ as $T$ increases ($k_B$ 
is the Boltzmann constant). These 
features are qualitatively similar to those of a trapped non-interacting 
condensate~\cite{Pethick, Napolitano}. However for an attractive condensate, 
there are important changes in the nature. This is due to the fact that the 
number of available energy levels of the system is limited, especially when 
$N \rightarrow N_{cr}$, while for any $N$, there are infinitely many energy levels for the repulsive or non-interacting condensate. In the limit of high $T$, both the 
repulsive and the non-interacting condensate behave as the corresponding trapped Bose-gas, resulting in a saturation in $C_N(T)$. On the other hand, an attractive condensate also shows similar behavior, only if it 
is allowed to absorb energy internally through rotational motion involving 
large orbital angular momenta. \\

We provide an understanding of this peculiar nature based on the many-body 
picture. For the theoretical calculation, 
we adopt the correlated potential harmonic (CPH) 
method~\cite{Das1,Das2} to solve the many-body problem approximately. This 
technique is based on the potential harmonics (PH) expansion method~\cite{Fabre}.  
The laboratory BEC must be very dilute to preclude three-body collisions, 
which lead to molecule formation and consequent depletion. Hence only 
two-body correlations are relevant. The PH is a subset~\cite{Fabre} of the full hyperspherical 
harmonics (HH) basis~\cite{Ballot}, that involves only two-body correlations. Hence 
the PH basis is a good approximation for expanding the condensate wave function. 
It reduces the bulk of the numerical procedure immensely, while retaining the most important 
basic features of the condensate. However, the leading members of the PH 
basis do not have the correct short separation behavior of the 
interacting Faddeev component. This causes a very slow rate of convergence of the PH 
expansion basis. To correct for this, we include a short-range correlation function 
in the expansion basis. This correlation function 
is obtained as the zero-energy solution of the two-body 
Schr\"odinger equation~\cite{Das2}. It is a correct representation of the 
short separation behavior and also incorporates the $s$-wave scattering length, 
$a_s$, through its asymptotic behavior~\cite{Pethick}. The technique 
has been shown to reproduce known results, both experimental and 
theoretical~\cite{PHEM}. These include the following: ground state 
properties (energy, wave function, condensate size, one-body density, 
pair-correlation, etc.) as also multipolar moments of both repulsive and 
attractive condensates, correct prediction of the critical number and 
collapse scenario of attractive condensates, 
thermodynamic properties of repulsive condensates, properties 
of condensates in finite traps, etc.\\


We can understand the behavior of heat capacity of attractive condensates in terms of the energy levels of the system produced by the CPH method. This method generates an effective potential in which the condensate moves. For an attractive condensate, the effective potential has a metastable region (MSR), separated from a deep well on the inner side by an intermediate finite barrier. A finite number of energy levels are supported by the MSR. As the number $N$ of atoms increases, the MSR shrinks and the number of energy levels reduce drastically. As temperature increases, particles are distributed in higher energy levels, according to Bose distribution formula. Thus at low temperatures the internal energy and $C_N(T)$ increase with temperature. At higher temperatures, the bosons have fewer levels to occupy, causing $C_N(T)$ to differ from the repulsive case. There is also a dominant effect of the drastically reducing number of energy levels as $N \rightarrow N_{cr}$.\\

The paper is organized as follows. For easy readability and to introduce our 
notations, we briefly review the 
correlated potential harmonic method in Section~II. Section~III provides 
our numerical procedure. Results and discussion are presented in 
Section~IV. Finally we draw our conclusions in Section~V. \\

\section{Correlated potential harmonics (CPH) method}

We adopt the correlated potential harmonics method~\cite{Das1,Das2} to 
solve the many-body problem of the BEC. We briefly recapitulate 
the technique in the following. Interested readers can find details 
in the cited references. \\

For the relative motion of a system of $N$ identical spinless bosons, we 
introduce $(N-1)$ Jacobi vectors
\begin{equation}
 \vec{\zeta}_{i}=\sqrt{\frac{2i}{i+1}}\left( \vec{x}_{i+1} - \frac{1}{i} 
\sum_{j=1}^{i} \vec{x}_{j}\right) , \hspace*{0.5cm} (i=1,...,N-1) ,
\end{equation}
where $\vec{x}_i$ is the position vector of the $i$-th particle. The 
Schr\"odinger equation governing the relative 
motion of the system trapped in a harmonic well, is 
\begin{eqnarray} 
\Big[&-&\frac{\hbar^{2}}{m} \sum_{i=1}^{\cal {N}} \nabla_{\vec{\zeta}_{i}}^{2}+
V_{trap}(\vec{\zeta}_1, ..., \vec{\zeta}_{\cal {N}}) + 
\nonumber
\\
&&V(\vec{\zeta}_{1}, ..., \vec{\zeta}_{\cal{N}})-E_{R} \Big] 
\psi(\vec{\zeta}_{1}, ..., \vec{\zeta}_{\cal{N}}) = 0 ,
\label{SE_rel_motion}
\end{eqnarray} 
where ${\cal{N}}=N-1$ and the trapping potential $V_{trap}$ and 
the interatomic interaction $V$ are 
expressed in terms of the Jacobi vectors. The energy of 
the relative motion is $E_R$. Next, we introduce hyperspherical 
variables corresponding to the set of ${\cal N}$ Jacobi vectors. 
First, a hyperradius is defined as 
\begin{equation}
r=\left[\sum_{i=1}^{\cal N} \zeta_i^2\right]^{\frac{1}{2}}.
\end{equation}
The remaining set of $(3{\cal N}-1)$ `hyperangles' consists of $2{\cal N}$ polar angles 
of ${\cal N}$ 
Jacobi vectors and $({\cal N}-1)$ angles defining their relative lengths~\cite{Ballot}. 
In the hyperspherical harmonics expansion method 
(HHEM) $\psi$ is expanded in the complete 
set of hyperspherical harmonics (HH), which are the eigenfunctions of 
the grand orbital operator [hyperangular part of 
the $\cal{N}$ dimensional Laplace operator, given by the sum in the 
first term of Eq. (\ref{SE_rel_motion})]~\cite{Ballot}. Substitution of this in 
Eq. (\ref{SE_rel_motion}) and 
projection on a particular HH result in a set of coupled differential 
equations. Imposition of symmetry of the wave function and calculation 
of the matrix elements become increasingly difficult and tedious as $N$ 
increases. In addition, the  
degeneracy of the HH basis increases very rapidly~\cite{Ballot} with the increase in the 
grand orbital quantum number $K$. Hence a convergent calculation using  
HHEM with {\it the full} HH basis is extremely computer intensive 
and unmanageable for $N > 3$. This is 
the price one pays for keeping {\it all many-body correlations} in $\psi$. \\

However all these complications can be avoided and a much simpler 
computational procedure can be formulated for the laboratory BEC, 
which is designed to be {\it extremely dilute} (typical number density 
is $\sim 10^{15}$ cm$^{-3}$) in order to 
avoid recombination through three-body collisions. Thus three-body 
correlations and three-body forces are totally negligible. 
We can then express $\psi$ as a sum of two-body Faddeev component 
$\psi_{ij}$ for the $(ij)$-interacting pair~\cite{Fabre}
\begin{equation}
\psi=\sum_{i,j>i}^N\psi_{ij}(\vec{r}_{ij},r).
\end{equation}
Note that the assuption of two-body correlations only makes 
$\psi_{ij}$ a function of the pair separation vector and the hyperradius only.
One can then expand $\psi_{ij}$ in a subset of HH, called the potential 
harmonics (PH) subset, which is sufficient for the expansion of 
the interaction potential $V(\vec{r}_{ij})$ as a function in the 
hyperangular space for the $(ij)$-partition. 
Since the labeling of the particles is arbitrary, we can choose 
$\vec{r}_{ij}=\vec{\zeta}_{\cal N}$. Then the corresponding PH, 
$\mathcal{P}_{2K+l}^{lm}(\Omega^{ij}_{\cal N})$ (the argument is 
the full set of hyperangles for the $(ij)$-partition) is 
independent of $\{\vec{\zeta}_1,\dots,\vec{\zeta}_{{\cal N}-1}\}$ 
and a simple analytic expression is possible~\cite{Fabre}. 
Expansion of the Faddeev component in the PH basis reads 
\begin{equation}
 \psi_{ij}(\vec{r}_{ij},r)=r^{-\frac{(3{\cal N}-1)}{2}}
\sum_{K}\mathcal{P}_{2K+l}^{lm}(\Omega^{ij}_{\cal N})u_{K}^{l}(r), 
\label{PHexpn}
\end{equation}
where $\mathcal{P}_{2K+l}^{lm}(\Omega^{ij}_{\cal N})$ is a potential 
harmonic~\cite{Fabre}.
The $r$-dependent factor in front is included to remove the first 
derivative with respect to $r$. Substitution of this expansion in 
the Faddeev equation for  the $(ij)$-partition 
\begin{equation}
 (T+V_{trap}-E_{R})\psi_{ij}=-V(r_{ij})\sum_{k,l>k}^{N} \psi_{kl},
\label{Faddeveq}
\end{equation}
[where $T=-\frac{\hbar^{2}}{m}\sum_{i=1}^{\cal N} \nabla^{2}_{\vec{\zeta}_{i}}$]
and projection on the PH corresponding to the $(ij)$-partition 
give a set of coupled differential equations in $r$. Note 
that any realistic two-body potential, $V(\vec{r}_{ij})$ can be 
used. A realistic interatomic potential has a very strong 
repulsion (arising from the nucleus-nucleus repulsion) at 
very short separations. Consequently, corresponding $\psi_{ij}$ 
must be vanishingly small for small values of $r_{ij}$. 
But the leading PH (corresponding to $K=0$) in the expansion in Eq. (\ref{PHexpn}) 
is a constant and 
does not have this behavior. Hence convergence of 
the expansion in Eq. (\ref{PHexpn}) will be very slow. To improve 
the rate of convergence, we include a short-range correlation function 
$\eta(\vec{r}_{ij})$ in the expansion basis, so that Eq. (\ref{PHexpn}) 
is replaced by 
\begin{equation}
 \psi_{ij}(\vec{r}_{ij},r)=r^{-\frac{(3{\cal N}-1)}{2}}
\sum_{K}\mathcal{P}_{2K+l}^{lm}(\Omega^{ij}_{\cal N})u_{K}^{l}(r)
\eta(\vec{r}_{ij}).
\label{CPHexpn}
\end{equation}
The short-range correlation function is chosen 
to have the expected behavior of $\psi_{ij}(\vec{r}_{ij},r)$ for 
small $r_{ij}$ in the following manner. The small $r_{ij}$ 
behavior of $\psi_{ij}$ will be that of a zero-energy pair 
interacting via $V(\vec{r}_{ij})$, 
since the energy of the interacting pair is practically zero. 
We obtain $\eta(\vec{r}_{ij})$ by solving the zero-energy two-body 
Schr\"odinger equation
\begin{equation}
-\frac{\hbar^2}{m}\frac{1}{r_{ij}^2}\frac{d}{dr_{ij}}\left(r_{ij}^2
\frac{d\eta(r_{ij})}{dr_{ij}}\right)+V(r_{ij})\eta(r_{ij})=0. 
\label{2Beqn} 
\end{equation}
Inclusion of the short-range correlation function $\eta(\vec{r}_{ij})$ 
enhances the rate of convergence greatly, which has been checked in our 
numerical calculation. \\

The laboratory BEC is very dilute; hence the average separation of 
the atoms is very large compared with the range of interatomic 
interactions. Moreover, the atoms scatter with almost zero energy. 
Hence the effective two-body interaction is represented by the 
$s$-wave scattering length $a_s$. In our calculation, we take 
$V(\vec{r}_{ij})$ to be the van der Waals potential with a hard 
core: $V(\vec{r}_{ij})=-\frac{C_6}{{r_{ij}}^6}$ for $r_{ij} \geq r_c$ and 
$=\infty$ for $r_{ij} < r_c$. The correlation function 
obtained by solving Eq. (\ref{2Beqn}) quickly attains its 
asymptotic form $C(1-\frac{a_{s}}{r_{ij}})$ for large $r_{ij}$. 
The asymptotic normalization is chosen to make the wavefunction 
positive at large $r_{ij}$. The hard core radius $r_c$ is 
adjusted so that the calculated $a_s$ is the actual experimental 
value of the scattering length~\cite{Pethick}. This procedure assures that the 
realistic two-body interaction appropriate for the condensate 
has been incorporated. \\

Substitution of the expansion, Eq.~(\ref{CPHexpn}) in 
Eq.~(\ref{Faddeveq}) and projection on the PH corresponding to the 
$(ij)$-partition result in 
\begin{eqnarray}
\Big[&-&\dfrac{\hbar^{2}}{m} \dfrac{d^{2}}{dr^{2}} + \dfrac{\hbar^{2}}{mr^{2}}
\{ {\cal L}({\cal L}+1) + 4K(K+\alpha+\beta+1)\} 
\nonumber 
\\
&+& V_{trap}(r) - E_R  \Big] U_{Kl}(r) \nonumber \\
&+& \sum_{K^{\prime}}f_{Kl}V_{KK^{\prime}}(r)f_{K'l} U_{K^{\prime}l}(r) = 0,
\label{CDE}
\end{eqnarray}
where $U_{Kl}(r) = f_{Kl}u_{K}^{l}(r)$,  ${\cal L} =
l+\frac{3N-6}{2}$, $\alpha=\frac{3N-8}{2}$, $\beta=l+\frac{1}{2}$, 
$l$ being the orbital angular momentum contributed by the 
interacting pair. 
$f_{Kl}^2$ is a constant representing the overlap of the PH for interacting
partition with the full set of all partitions, which can be 
found in Ref.~\cite{Fabre}. The correlated potential matrix 
element $V_{KK^{\prime}}(r)$ is given by~\cite{Das2} 
\begin{eqnarray}
V_{KK^{\prime}}(r) &=& (h_{K}^{\alpha\beta}
h_{K^{\prime}}^{\alpha\beta})^{-\frac{1}{2}}
\int_{-1}^{+1} \Big\{ 
P_{K}^{\alpha 
\beta}(z)
V\left(r\sqrt{\frac{1+z}{2}}\right) 
\nonumber 
\\
&&P_{K^{\prime}}^{\alpha \beta}(z)\eta\left(r\sqrt{\frac{1+z}{2}}\right)
W_{l}(z) \Big\} dz.
\label{corrPME}
\end{eqnarray}
Here $h_{K}^{\alpha\beta}$ and $W_{l}(z)$ are respectively the norm
and weight function~\cite{Abramowitz} of the Jacobi polynomial
$P_{K}^{\alpha \beta}(z)$. 
Note that the inclusion of the short-range correlation function, 
$\eta(r_{ij})$ makes the PH basis non-orthogonal. Numerical 
solution of Eq. (\ref{2Beqn})  shows that  
$\eta(r_{ij})$ differs from a constant value only in a small interval 
of small $r_{ij}$ values. Hence the dependence of the overlap 
$<{\mathcal P}^{lm}_{2K+l}(\Omega^{(ij)}_N)|
{\mathcal P}^{lm}_{2K+l}(\Omega^{(kl)}_N)\eta(r_{kl})>$ on the 
hyperradius $r$ is quite small. Disregarding 
derivatives of this overlap with respect to the hyperradius, 
we approximately get Eq. (\ref{CDE}), 
with $V_{KK^{\prime}}(r)$ given by Eq. (\ref{corrPME}). The effect 
of the overlap being different from 
unity is represented by the asymptotic constant $C$ of $\eta(r_{ij})$. 
The emerging physical picture is: the effective interaction between pairs 
of atoms at very low energy becomes $V(r_{ij})\eta(r_{ij})$. This is 
justified, since at very low kinetic energy, the atoms have a very large 
de Broglie wave length and do not approach each other close enough to 
"see" the {\it actual} interatomic interaction. In the limit of 
zero energy, the scattering cross section 
becomes $4\pi|a_s|^2$ and the effective interaction is governed by 
the $s$-wave scattering length $a_s$, through the asymptotic form of 
$\eta(r_{ij})$. \\

Introduction of the PH basis and inclusion of the short-range 
correlation function, referred to as the correlated potential 
harmonic (CPH) method, simplifies the many-body problem dramatically. 
A fairly fast computer code can solve Eq. (\ref{CDE}) with upto 
15000 particles in the condensate. This technique has been 
tested against known results, both experimental ones and theoretical 
ones calculated by other authors, 
for repulsive as well as attractive condensates~\cite{Das1,Das2,PHEM}.\\

\section{Numerical procedure}

\subsection{Solution of coupled equations}

Although Eq. (\ref{CDE}) can be solved by an exact numerical technique 
using the Numerov method, we adopt the hyperspherical adiabatic approximation 
(HAA)~\cite{das3}, which apart from simplifying the computations 
greatly, provides an effective potential in the hyperradial 
space, in which the condensate moves. This effective condensate 
potential provides a physical picture for the internal mechanism 
of the condensate.\\

In the HAA, one assumes that the hyperangular motion 
is much faster than the hyperradial motion, since the latter corresponds to 
the breathing mode. Therefore, one can 
solve the former adiabatically for a fixed value of $r$ and 
obtain the solution as an effective potential for the hyperradial 
motion, as in Born-Oppenheimer approximation. The hyperangular motion is solved by diagonalizing the potential 
matrix $V_{KK^{\prime}}(r)$ together with the hyper-centrifugal 
potential [second term of Eq. (\ref{CDE})]. 
The lowest eigenvalue $\omega_0(r)$ [corresponding 
eigen column vector being $\chi_{K0}(r)$], is the effective potential 
for the hyperradial motion~\cite{das3}: 
\begin{equation}
\Big[-\dfrac{\hbar^2}{m}\dfrac{d^2}{dr^2}+\omega_0(r)+
\sum_K|\dfrac{\chi_{K0}(r)}{dr}|^2-E_R\Big]\zeta_0(r)=0.
\label{HAAeq}
\end{equation}
The third term is an overbinding correction. 
Eq. (\ref{HAAeq}) is solved by the Runga-Kutta method, subject 
to appropriate boundary conditions to get $E_R$ 
and the hyperradial wave function $\zeta_0(r)$. The many-body wave 
function can be constructed in terms of $\zeta_0(r)$ and 
$\chi_{K0}(r)$~\cite{das3}. Total energy is obtained by adding the 
center of mass energy ($1.5$ o.u.) to $E_R$. Energy levels, $E_{nl}$, are 
characterized by the quantum numbers $(n,l)$, where $n$ represents 
the excitation quantum number for a given orbital angular momentum 
$l$. The HAA has been tested for nuclear, atomic and molecular systems 
and shown to give better than 1\% accuracy, even for the long-range 
Coulomb potential~\cite{HAAaccu}. In our case, the van der Waals 
potential has a shorter range and HAA is expected to be better. Moreover, 
in a BEC, the dominant confining harmonic oscillator potential is 
smooth and the corresponding hyperradial equation is completely 
decoupled. Hence in a BEC, the HAA is expected to be far better. 
We tested this by solving the CDE, Eq.~(\ref{CDE}), with the interatomic potential for 
the ground state by the renormalized Numerov method~\cite{Ghosh,Johnson}, 
which is an exact numerical algorithm for solving a set of coupled 
differential equations. The calculated exact ground state energies 
are (in o.u.) $948.6420, 1174.5284, 1277.8219, 1460.3706$ and $1596.2611$ 
respectively for $N=700, 900, 1000, 1200$ and $1400$. These compare very well 
with the corresponding HAA results: $948.0986, 1173.9809, 1277.2040, 
1459.6373$ and $1595.8163$ respectively. The error is less than 
$0.06$\% in all cases. Thus we can safely use the HAA, which reduce 
the numerical complications to a great extent. \\

\subsection{Calculation of specific heat}

At a temperature $T>0$, bosons are distributed in available energy levels 
$E_{nl}$ according to Bose distribution function 
\begin{equation}
 f(E_{nl}) = \frac{1}{e^{\beta(E_{nl}-\mu)} -1}
\label{distribution}
\end{equation}
where $\beta = 1/{k_{B}}T$ and $\mu$ is the chemical potential. The latter 
is determined 
from the constraint that the total number of particles is $N$. Clearly, $\mu$  
has a temperature dependence. 
The total number of bosons in the trap is fixed and at any temperature 
it can be written as 
\begin{equation}
 N = \sum_{n=0}^{\infty} \sum_{l=0}^{\infty} (2l+1)f(E_{nl})
 \label{Nsum}
\end{equation}
At a particular temperature $T$, $\mu$ is determined from the 
constraint Eq. (\ref{Nsum}). 
The total energy of the system at $T$ is given by 
\begin{equation}
 E(N,T)=\sum_{n=0}^{\infty} \sum_{l=0}^{\infty}(2l+1)f(E_{nl})E_{nl}
 \label{Esum}
\end{equation}
The specific heat at fixed particle number $N$ is calculated using 
the relation
\begin{equation}
 C_{N}(T) = \frac{\partial E(N,T)}{\partial T}{\Big |}_{N}
\label{spheat}
\end{equation}
Using (\ref{distribution}), (\ref{Esum}), (\ref{spheat}) one can obtain the 
heat capacity as 
\begin{eqnarray}
C_{N}(T) = \beta\sum_{n=0}^{\infty}\sum_{l_n=0}^{\infty} \frac{(2l_n+1)E_{nl_n}\exp{(\beta(E_{nl_n}-\mu))}}{(\exp{(\beta(E_{nl_n}-\mu))}-1)^2} \nonumber \\
 \times {\Big[}\frac{E_{nl_n}-\mu}{T}+\frac{\partial \mu}{\partial T}{\Big]} \nonumber \\
\label{spheatform}
\end{eqnarray}
where
{\tiny{
\begin{eqnarray}
\frac{\partial \mu}{\partial T} = \hspace*{7cm}\nonumber \\
  - \frac{\sum_{m=0}^{\infty}\sum_{l_m=0}^{\infty}(2l_m+1)(E_{ml_m}-\mu)\exp{(\beta(E_{ml_m}-\mu))}
(f(E_{ml_m}))^2}{T\sum_{p=0}^{\infty}\sum_{l_p=0}^{\infty}(2l_p+1)\exp{(\beta(E_{pl_p}-\mu))}(f(E_{pl_p}))^2}
\end{eqnarray} 
}}
For an ideal non-interacting bosonic gas containing $N$ 
bosons in a three-dimensional 
isotropic harmonic well, the critical temperature 
$T_{c}^0$ is well defined~\cite{Dalfovo}. $\mu$ remains equal to the 
energy of the single particle ground state for 
$T<T_c^0$ and start decreasing rapidly for $T>T_c^0$. In the 
standard text book treatment~\cite{Huang} the sums in 
eqs. (\ref{Nsum}-\ref{Esum}) are replaced in the semi-classical approximation 
by integrals over energy, assuming a continuous energy spectrum. In a harmonic trap 
the energy spectrum is discrete and this assumption is not valid, particularly 
for small $N$ at low energies. A correct treatment~\cite{Napolitano} shows that $\mu$ 
{\it decreases slowly} from its maximum value (equal to the ground state energy) 
as $T$ increases from zero, the rate of decrease becoming suddenly 
rapid at some temperature close to the reference temperature $T_c^0$ corresponding 
to the same value of $N$~\cite{Biswas1}. Thus, in this case the critical 
temperature is not well defined. In the correct treatment, the 
heat capacity also becomes a smooth function of $T$, attaining a maximum at 
a temperature, at which $\mu$ suddenly becomes a rapidly 
decreasing function of $T$. In the limit of large $N$, the behaviors of $\mu(T)$ 
and $C_N(T)$ curves approach those of the text book treatment. 
The {\it transition temperature} $T_c$ for a finite interacting 
system is defined as the temperature at which 
$C_N(T)$ is a maximum~\cite{Napolitano}
\begin{equation}
\frac{\partial C_N(T)}{\partial T}{\Big|}_{T_c}=0
\label{defTc}
\end{equation}

For the numerical calculation for a chosen particle number 
$N$, the CPH equations are solved for 
a large number of energy levels -- typically $n$ running from 
$0$ to $300$ and $l$ running from $0$ to $200$, subject to an 
upper energy cutoff value, $E_{UL}$ (see below), so that 
$E_{nl}<E_{UL}$. Using previously calculated values of 
$E_{nl}$, Eq. (\ref{Nsum}) is solved for $\mu$, at a chosen 
temperature $T$, by a 
modified bisection method. Next $E_{UL}$ is increased and 
the process repeated, until convergence in $\mu$ is achieved. 
Using this upper energy cutoff, $C_N(T)$ is determined using 
Eq.~(\ref{spheatform}). \\

\section{Results and discussions}

\begin{figure}[hbpt]
\vspace{-5pt}
\centerline{
\hspace{-3.3mm}
\rotatebox{270}{\epsfxsize=6.5cm\epsfbox{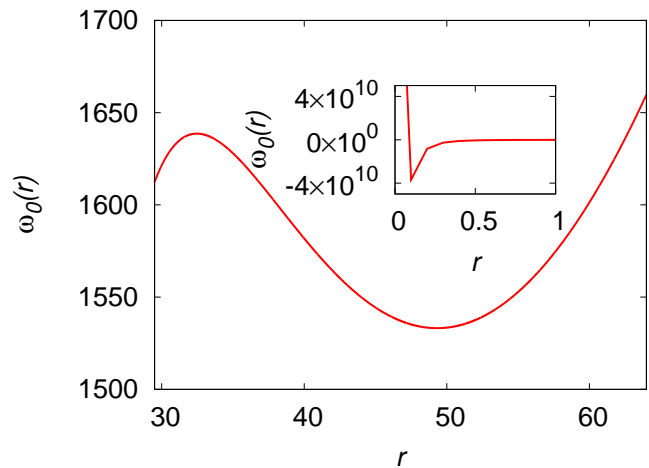}}} 
\caption{(Color online)
Calculated effective potential $\omega_0(r)$ against $r$ in o.u.
for the attractive $^7$Li condensate with $N=1300$. The narrow 
and deep well near origin is shown in the inset (note that different 
scales are used).} 
\label{w0_r}
\end{figure}

We consider the attractive condensate of $^7$Li atoms in the 
trap used in the experiment at Rice University~\cite{Li-Bradley}. The 
harmonic trap used was axially symmetric with 
$\nu_x=\nu_y=163$ Hz and $\nu_z=117$ Hz. For 
simplicity, we consider an isotropic trap with $\nu=
(\nu_x\nu_y\nu_z)^{\frac{1}{3}}$. The experimental 
value of $a_s$ is $-27.3$ o.u. We use oscillator unit (o.u.) 
of length ($\sqrt{\frac{\hbar}{m\omega}}$) and energy 
($\hbar\omega$). As mentioned earlier, we choose van der Waals (vdW) 
potential for the interatomic interaction, with 
known value~\cite{Pethick} of 
$C_6=1.715\times10^{-12}$ o.u. The value of $r_c$ is obtained by the procedure 
discussed following  
Eq. (\ref{2Beqn}), so that calculated $a_s$ has the experimental 
value~\cite{Pethick}. Its numerical value is $5.338 \times 10^{-4}$ o.u. 
The calculated effective potential, $\omega_0(r)$ 
is plotted as a function of $r$ in Fig. \ref{w0_r} for $N=1300$. 
In the $r \rightarrow 0$ limit, 
$\omega_0(r)$ becomes strongly repulsive, due to the repulsive core 
of vdW potential and the hypercentrifugal repulsion of Eq. (\ref{CDE}). 
As $r$ increases, there is a deep narrow well (DNW) arising from 
the strong interatomic attraction at small values of $r$. This 
attraction is proportional to the number of pairs and hence 
increases rapidly as $N$ increases. For 
still larger $r$, the effects of the kinetic pressure 
(including the centrifugal repulsion), interatomic attraction and the 
harmonic confinement together produce a metastable region (MSR). An 
intermediate barrier (IB) appears between the DNW and MSR. The 
DNW is very deep and narrow, hence it is shown as an inset in Fig. \ref{w0_r} 
(note large changes in scale for both horizontal and 
vertical axes). 
\begin{figure}[hbpt]
\vspace{-5pt}
\centerline{
\rotatebox{0}{\epsfxsize=8.0cm\epsfbox{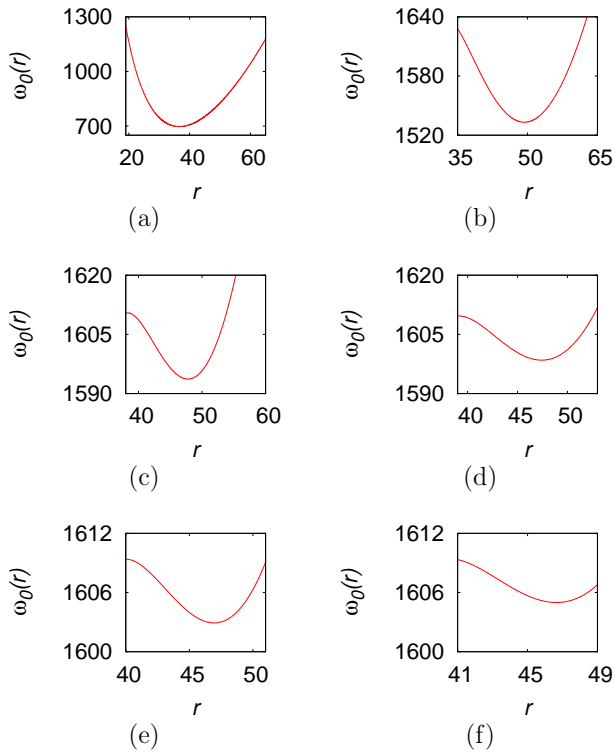}}} 
\caption{(Color online)
Plot of effective potential $\omega_0(r)$ against $r$ (both in 
appropriate o.u.) 
for the attractive $^7$Li condensate with $N=500$, $1300$, 
$1400$, $1410$, $1420$ and $1426$ in panels (a) -- (f) 
respectively, showing how the MSR shrinks in depth and width. 
Note that different scales have been used in different panels, 
to bring out the features of the MSR as $N \rightarrow N_{cr}$.} 
\label{MSR}
\end{figure}
As 
$N$ increases, the DNW becomes deeper, IB shallower and the minimum 
of the MSR higher. At the critical value $N_{cr}$, the maximum of IB and 
the minimum of MSR merge to form a point of inflexion and the MSR 
disappears. At this point, the condensate falls into the DNW, 
resulting in a collapse of the condensate and formation of clusters 
within the DNW. 
Our calculated value of $N_{cr}$ is 1430. In panels (a) -- (f) of 
Fig. \ref{MSR}, we demonstrate how the MSR shrinks, with $N$ approaching 
$N_{cr}$, for $N=500, 1300,1400,1410,1420,$ and $1426$ respectively. 
From Fig. \ref{MSR}, one notices that both the depth and width of MSR 
decrease as $N$ increases towards $N_{cr}$. Hence the number of 
bound energy levels supported by the MSR decreases rapidly with 
$N$ (see also Fig. \ref{en_levels}). \\

\begin{figure}[hbpt]
\vspace{-5pt}
\centerline{
\hspace{-3mm}
\rotatebox{270}{\epsfxsize=6.5cm\epsfbox{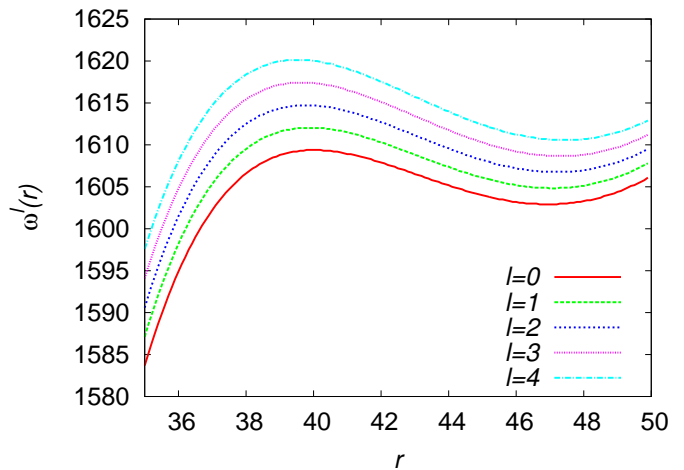}}}
\caption{(Color online)
Plot of effective potential $\omega^{l}(r)$ against $r$ (both in 
appropriate o.u.) 
for the attractive $^7$Li condensate with $N=1420$ atoms, 
corresponding to $l=0,1,2,3,4$. The curves show how 
the IB increases as $l$ increases.}
\label{w_all}
\end{figure}

\begin{figure}[hbpt]
\vspace{-5pt}
\centerline{
\hspace{1mm}
\rotatebox{270}{\epsfxsize=6.5cm\epsfbox{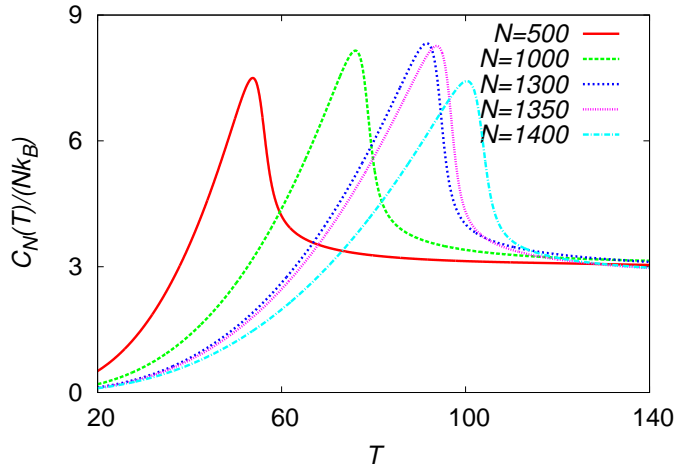}}} 
\caption{(Color online)
Plot of heat capacity $C_N(T)/(Nk_B)$ (dimensionless) against 
$T$ (in nK), for indicated number of $^7$Li atoms in 
the metastable condensate.}
\label{CN_N}
\end{figure}

However, 
the effective potentials shown in Fig. \ref{w0_r} and Fig. \ref{MSR} are obtained for 
$l=0$. For higher $l$, the effective potential has a higher IB, arising 
from the $l$-dependent terms of the hyper-centrifugal repulsion 
[see Eq. (\ref{CDE})]. Thus the position of the MSR rises higher in energy 
as $l$ increases, as can be seen in Fig. \ref{w_all} for $l=0,1,2,3,4$ for a 
condensate containing 1420 atoms. 
Hence particles with $l>0$ can attain higher energy levels. 
Inclusion of these levels will have a profound effect on the 
heat capacity. If 
such energy states were ignored ({\it i.e.}, only the energy levels 
supported by the $l=0$ MSR considered), the heat capacity would reduce 
drastically and $T_c$ would increase indefinitely as 
$N \rightarrow N_{cr}$, since all the atoms would be forced into the 
few remaining energy levels available for internal excitation. In our calculation, we have 
retained all energy levels supported by a given $l$. A question 
arises as to whether the metastable condensate can have large 
$l$ values. Intuition indicates that, with increase of temperature, 
the system can absorb energy only by increasing its rotational 
kinetic energy, thereby increasing the stability of the metastable 
system with enhanced centrifugal repulsion. Increase of kinetic 
energy due to faster linear motion alone 
would cause the system to fall in the DNW near the center of the 
condensate. Compared with the 
non-interacting or repulsive condensates, the attractive condensate 
has a clear distinction, {\it viz.} while the number of hyperradial 
excitations for a given $l$ in the former is not limited, it is 
drastically limited in the latter. Thus for an attractive condensate, 
there are fewer energy states, in which the system can reside. 
This causes $(C_N)_{max}$ to increase initially for $N \ll N_{cr}$ 
(when energy levels are not greatly restricted), but as $N \rightarrow 
N_{cr}$, it starts decreasing, after attaining a maximum. Fig. \ref{CN_N} 
shows how $C_N(T)$ depends on $T$, for selected values of $N$. 
It is seen that the transition temperature $T_c$ increases gradually 
with $N$, but $(C_N)_{max}$ increases up to $N=1300$ and for larger 
$N$, it starts decreasing. 
When $N \ll N_{cr}$, the nature is similar to that of a repulsive 
condensate~\cite{Biswas1}, since in this case, the number of available energy 
levels are still large enough (the top most energy level -- including $l \neq 0$ -- in the MSR 
has an energy much greater than $k_BT$), 
so that the top most levels are still
practically unoccupied and 
there is scope for further internal excitation 
as $T$ increases. 
Consequently $T_c$ increases gradually with $N$, 
as in the repulsive case. As $N$ approaches $N_{cr}$, the number of 
energy levels supported by the MSR 
decreases rapidly and there is less scope for absorbing energy 
internally as $T$ increases. Hence $(C_N)_{max}$ 
decreases and $T_c$ increases faster, as $N$ increases 
towards $N_{cr}$. At higher temperatures, higher $l$ states are excited, 
which push atoms further outwards, increasing the average interatomic 
separation. Consequently, the system behaves ultimately as a 
non-interacting Bose gas. Thus the asymptotic value of $C_N(T)$ 
becomes $3Nk_B$. 
In Fig. \ref{CN_N}, we plot the dimensionless quantity $C_N(T)/(Nk_B)$ against $T$ (in nK) for $^7$Li condensate 
with $N=500, 1000, 1300, 1350$ and $1400$. The features discussed above are 
clearly visible. One notices that the behavior for $N < 1300$ is 
similar to that of a repulsive condensate, but 
as $N$ exceeds $1300$, the curves become 
\begin{figure}[hbpt]
\vspace{-10pt}
\centerline{
\hspace{-3.3mm}
\rotatebox{270}{\epsfxsize=6.5cm\epsfbox{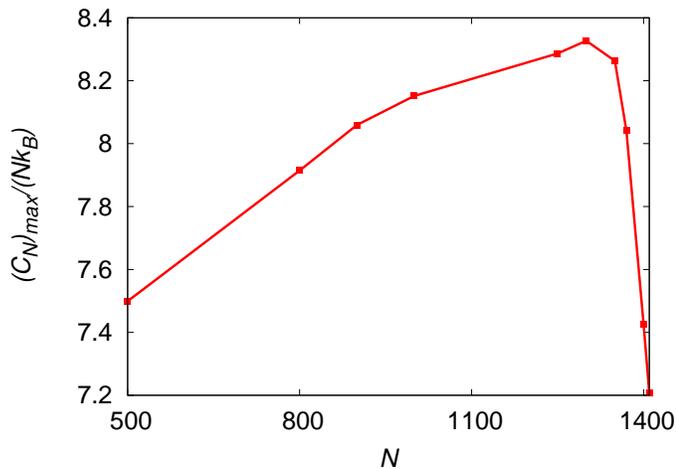}}}
\caption{(Color online)
Plot of the peak heat capacity, $(C_N)_{max}/(Nk_B)$ (dimensionless), against 
the number $N$ of bosons in the attractive $^7$Li 
condensate.}
\label{CNmax} 
\end{figure}
flatter near their 
maxima and the peak value of $C_N(T)/(Nk_B)$, namely, 
$(C_N)_{max}/(Nk_B)$, decreases fairly rapidly, as $N \rightarrow N_{cr}$. 
All the curves appear to converge to the Bose gas limit. But a closer 
scrutiny shows that the curves for $N=1350$ and $1400$ show a slight downward trend. This is due to a limitation in the higher energy cut-off 
used in our calculation. 
In Fig. \ref{CNmax}, we plot calculated $(C_N)_{max}/(Nk_B)$ as a function of $N$. 
It is seen that this quantity increases gradually up to $N=1300$. 
Beyond this value, $(C_N)_{max}/(Nk_B)$ decreases fairly rapidly 
as $N \rightarrow N_{cr}$. 
\begin{figure}[hbpt]
\vspace{-1pt}
\centerline{
\hspace{-3.3mm}
\rotatebox{270}{\epsfxsize=6.5cm\epsfbox{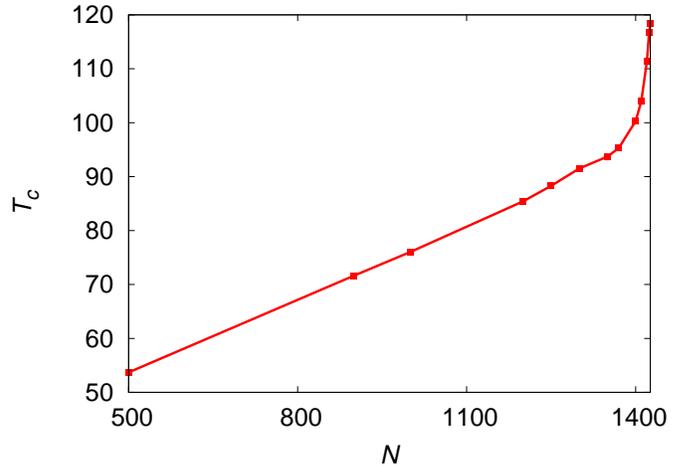}}}
\caption{(Color online)
Plot of transition temperature (in nK) versus $N$ for the attractive $^7$Li 
condensate.} 
\label{Tc}
\end{figure}
A plot of transition temperature $T_c$ (in nK) as a function of $N$ 
is shown in Fig. \ref{Tc}. Initially $T_c$ increases linearly for 
$N < 1300$. As discussed above, this behavior is expected for small 
$N$, as in the case of a repulsive condensate. But for $N > 1300$, $T_c$ 
increases rapidly. Both the decrease of $(C_N)_{max}$ and faster increase 
of $T_c$ are due to reduction 
\begin{figure}[hbpt]
\vspace{-5pt}
\centerline{
\hspace{-3.3mm}
\rotatebox{270}{\epsfxsize=6.5cm\epsfbox{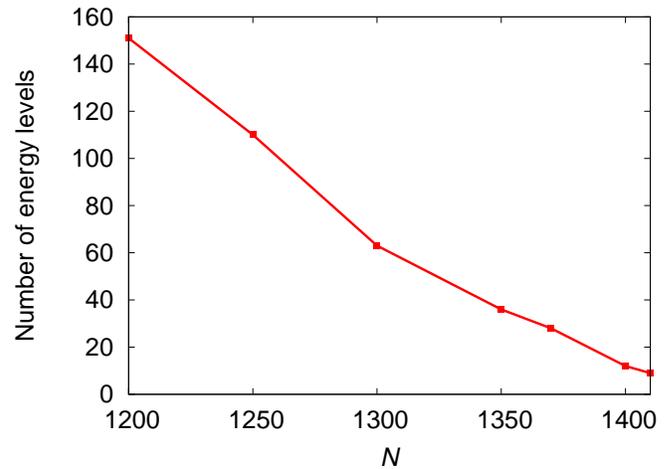}}} 
\caption{(Color online)
Plot of total number of available $l=0$ energy levels versus $N$, 
close to $N_{cr}$ for the attractive $^7$Li 
condensate.} 
\label{en_levels}
\end{figure}
in the number of available energy levels as 
$N$ approaches $N_{cr}$. We demonstrate this in Fig. \ref{en_levels} for the $l=0$ 
energy levels as $N$ increases from $1200$ to $N_{cr}$. 
The decrease in the number of available energy levels forces a larger 
fraction of the bosons to be in the ground state as T increases 
[see Eq. (\ref{Nsum})]. This causes a decrease of $(C_N)_{max}$ and an increase of $T_c$, as $N \rightarrow N_{cr}$. \\

A possible scenario of the attractive condensate as its temperature 
is gradually increased is the following. The standard definition of 
critical number $N_{cr}$ referred to in the literature, 
corresponds to the $l=0$ condensate at zero temperature. 
As $T$ is gradually raised, the system absorbs energy by occupying higher 
available energy levels upto the top of the MSR. However, there is a 
finite life time of atoms in higher energy levels due to tunneling through 
the IB into the DNW. Thus there will be a decrease in the number of atoms 
in the MSR. The rate of loss of atoms will increase with the energy of 
the level, as also with $N$ (increase of $N$ will lower the IB). 
The usual definition of heat capacity, 
$C_N(T)$, is the rate of change of internal energy with respect to $T$ 
[Eq. (\ref{spheat})], 
{\it for a fixed number $N$ of atoms in the condensate}. This definition 
is unambiguous for a repulsive or non-interacting condensate, since 
in these cases there is no loss. However, since the loss is appreciable 
for an attractive condensate for $N$ close to $N_{cr}$, or if $T$ is 
such that $k_BT$ is comparable with the highest excitation energy 
allowed by the IB, this definition 
demands that atoms be pumped into the condensate at the same rate as the 
loss rate from the condensate. Such an experimental procedure has not 
been adopted yet. However, for $N \ll N_{cr}$ and $T \ll T_c$, the highest 
appreciably occupied levels will have negligible tunneling probability 
into the DNW. Under these conditions, the metastable attractive condensate 
is fairly long lived and the standard definition of $C_N(T)$ is acceptable. 
Hence standard experimental techniques can be adopted. Thus our results 
presented in Figs.~4 -- 6 are experimentally verifiable in the small $N$, 
small $T$ limit. We have presented, for theoretical completeness, results 
for $N$ close to $N_{cr}$ and for $T$ beyond $T_c$. The question of how 
the system can absorb energy internally for such values of $N$ and $T$ was 
already discussed above. 

\section{Conclusions}
In this work, we report a detailed calculation of the heat capacity 
$C_N(T)$ of an attractive Bose-Einstein condensate containing $N$ 
atoms of $^7$Li. 
The correlated potential harmonics method, which is appropriate for the 
dilute BEC, has been used. 
The effective potential, in general, supports a large number of energy levels. 
At $T=0$, the lowest energy level accommodates all the bosons. As 
temperature increases, particles are distributed in higher energy levels, 
according to Bose distribution formula. Thus the internal energy of the 
system increases. Heat capacity $C_N$ for a fixed number $N$ of particles 
in the condensate is defined as the temperature derivative of the total 
internal energy. For a repulsive condensate trapped by an ideal harmonic 
oscillator, the effective potential has no upper cut off. Hence the energy 
levels are not limited in energy. Consequently, total internal energy and  
$C_N$ increase as temperature increases upto $T_c$. For $T > T_c$, the 
ground state occupation becomes suddenly microscopic (negligible) and the 
system behaves like a harmonically trapped Bose gas. 
Hence $C_N$ decreases rapidly above $T_c$, reaching its asymptotic value 
$3Nk_B$. Thus $C_N$ first increases, reaches a maximum value $(C_N)_{max}$ 
and then decreases  rapidly to its asymptotic value, as $T$ increases from zero. \\

For $N< N_{cr}$ bosons with mutual attraction, a metastable condensate is 
formed in the metastable region (MSR) of the effective potential. On the 
left of the MSR, an intermediate barrier (IB), followed by a deep narrow 
well and finally a strongly repulsive core appear, as one approaches 
the center of the condensate. For $N \ll N_{cr}$, the IB is very high  and 
the minimum of the MSR is very low, so that the metastable well is 
sufficiently deep compared with thermal excitation energy $k_BT_c$ at $T_c$, 
and a large number of energy levels are supported. Hence for $T \leq T_c$, 
even the most thermally excited particles do not feel the effect of the IB 
and $C_N$ increases gradually, as in the repulsive case. \\

With increase of temperature, the system with a fixed $N$ 
absorbs energy internally by increasing the occupation probability of 
higher energy levels supported by the metastable region of the 
effective potential. Atoms in energy levels close to the top of the 
intermediate barrier have appreciable probability to tunnel into 
the deep narrow well, causing the condensate to loose atoms. But such 
levels are not occupied with any appreciable probability if 
$N \ll N_{cr}$ and $T \ll T_c$. Hence such a condensate is quasi-stable 
and $C_N(T)$ calculated for a fixed $N$ is appropriate. 
When the rate of atom loss from the condensate is appreciable, 
standard definition of heat capacity at constant $N$ requires feeding the attractive 
condensate with additional atoms at a rate such as to compensate 
for the loss rate. Although this is not the usual experimental 
technique, we investigate such cases for a complete theoretical 
study. In such a situation, 
there are drastic changes. The peak value of $C_N(T)$ (the 
temperature at which this occurs is the transition temperature $T_c$) 
initially increases gradually with $N$, then after reaching a 
maximum, decreases fairly rapidly near $N_{cr}$. On the other hand, for 
small $N$, $T_c$ increases almost linearly up to $N \sim 1300$. 
For larger $N$, the transition temperature increases rapidly with 
$N$. We provide an explanation of this 
behavior, based on the microscopic mechanism of absorption 
of internal energy, as $T$ increases. As $N$ increases 
towards $N_{cr}$, the depth and width of the metastable well decrease 
rapidly. As a result, the number of 
energy levels supported by the metastable well decreases rapidly. 
This tends to increase $T_c$, since fewer energy levels are available 
for absorption of internal energy, and bosons are forced to be in lower 
energy levels as $T$ increases. Rapid reduction of available 
energy levels as $N \rightarrow N_{cr}$, causes quicker 
saturation of internal energy of the condensate. Consequently, 
the maximum of $C_N(T)$ decreases rapidly as $N \rightarrow N_{cr}$.

\section {Acknowledgement}
We would like to thank Dr. Parongama Sen for useful discussions. 
SG acknowledges CSIR (India) for a Junior 
Research Fellowship [Grant no. 09/028(0762)/2010-EMR-I], 
TKD acknowledges UGC (India) for the Emeritus 
Fellowship [Grant no. F.6-51(SC)/2009(SA-II)] and AB acknowledges CSIR (India) for a Senior 
Research Fellowship[Grant no. 09/028(0773)/2010-EMR-I].

\end{document}